\documentclass[12pt]{iopart}

\usepackage{graphicx}
\usepackage[usenames]{color}

\usepackage[usenames]{color}
\usepackage[normalem]{ulem}

\newcommand{\beq}{\begin{equation}}
\newcommand{\eeq}{\end{equation}} 
\newcommand{\beqa}{\begin{eqnarray}}
\newcommand{\eeqa}{\end{eqnarray}} 
\begin{document}

%\begin{multicols}{2}

\title[Bright  and dark-in-bright dipolar Bose-Einstein condensate solitons]{Two-dimensional bright  and dark-in-bright dipolar Bose-Einstein condensate solitons on a one-dimensional optical lattice}

\author{ S. K. Adhikari \footnote{adhikari@ift.unesp.br; URL: http://www.ift.unesp.br/users/adhikari}
} 
\address{
Instituto de F\'{\i}sica Te\'orica, UNESP - Universidade Estadual Paulista, 01.140-070 S\~ao Paulo, S\~ao Paulo, Brazil
} 

\begin{abstract}

We study the statics and dynamics of  anisotropic, stable,   bright and dark-in-bright
dipolar   quasi-two-dimensional Bose-Einstein condensate (BEC) solitons
on a one-dimensional (1D) optical-lattice (OL) potential. These solitons    mobile in a plane perpendicular to the 1D 
OL trap can have both repulsive and attractive contact interactions.  
The dark-in-bright solitons are the excited states of the bright solitons.
 The solitons, when subject to a small perturbation, exhibit sustained breathing oscillation. The dark-in-bright solitons can be created by phase imprinting a 
bright soliton. At medium velocities the collision between two solitons is found to be quasi
elastic.
 The results are demonstrated by a numerical simulation 
of the three-dimensional mean-field Gross-Pitaevskii 
equation   
in three 
spatial  dimensions employing realistic   interaction parameters for a dipolar $^{164}$Dy BEC.
%$^{164}$Dy BEC and a binary $^{164}$Dy-$^{162}$Dy BEC.

\end{abstract}


\maketitle

\section{Introduction}
 
A bright soliton can travel with a constant velocity   in one dimension (1D)
  maintaining its shape due to a cancellation of nonlinear attraction and dispersive
repulsion.  
    Solitons   have been studied in  water wave, Bose-Einstein condensates (BEC) \cite{gammal} and  nonlinear optics \cite{rmp}. Of these, solitons in BEC have drawn much attention lately
because of its inherrent quantum interaction. Quasi-1D 
bright matter-wave solitons  were predicted \cite{4} for {\it attractive} atomic interaction   and  created in  BECs of
$^7$Li \cite{1a,1b}
 and
$^{85}$Rb \cite{3}  atoms  by adjusting the atomic contact 
interaction to a suitable 
attractive value using a Feshbach resonance   \cite{fesh}. { Due to collapse instability
for attractive contact interaction in alkali-metal-atom BECs, (a)
one cannot have a quasi-two-dimensional (quasi-2D) bright soliton for a cubic   nonlinearity, and (b) the usual quasi-1D
bright solitons can accommodate a small number of atoms \cite{1a,3}.  }

The recent observation  of dipolar  BECs of $^{52}$Cr \cite{crex},   $^{164}$Dy \cite{ExpDy} and   $^{168}$Er \cite{ExpEr}  
 atoms  { have  started a great deal of activity in this area 
\cite{dbec1,dbec2,dbec3,dbec4,dbec5,dbec6,dbec7,dbec8,dbec9,dbec10}}
 including   new investigation of  BEC solitons in a different scenario.
Apart from quasi-1D bright dipolar   BEC   solitons  \cite{1DA},
asymmetric  quasi-2D bright  \cite{2D1,2D2,2D3,2D4,2dcol,2D6,2D7} and  vortex   \cite{ol2D1,ol2D2}
  solitons 
  have been predicted. Also, 
in a dipolar BEC, unlike in a nondipolar BEC, these solitons can be realized for a repulsive 
contact interaction. Consequently, one can have a
robust dipolar soliton which can accommodate a large number of atoms.   
There have been studies of collision of quasi-1D \cite{1DA,1dcol} and quasi-2D
 \cite{2dcol,ol2D1,2D2} dipolar solitons.

% {Although a spatial  optical quasi-2D soliton has been experimentally observed in  liquid carbon disulfide  with cubic-quintic nonlinearity \cite{cid}, a quasi-2D dipolar soliton has not yet been observed. }

  In this paper we study the statics and dynamics  of quasi-2D dipolar
bright and dark-in-bright
 solitons mobile in the $x-z$ plane and  trapped along the $y$ axis
by a  1D periodic optical lattice (OL)   potential.
The dipolar atoms will always 
be considered polarized along the $z$ direction. {
 A periodic OL potential cannot localize 
a BEC to form a bright soliton with only repulsive contact interaction, although, localized gap solitons in the band-gap of a periodic OL potential can be formed in a repulsive 
nondipolar \cite{gap1} and dipolar \cite{gap2} BEC.  The 
gap solitons are trapped in all  directions and hence
cannot move freely 
like a bright soliton without trap in certain direction(s). 
  Because of the anisotropic nature 
of the dipolar interaction and the OL potential, the bright and dark-in-bright
  quasi-2D dipolar solitons, which we consider here,
have a fully anisotropic shape.
For a small strength of the OL potential,  the  quasi-2D soliton extends over several OL sites 
in the $y$ direction. However, for a moderate strength of the OL potential and for a moderate 
number of dipolar atoms, compact quasi-2D solitons occupying a single OL site can be formed and we will be mostly interested in such compact solitons. For a large number of dipolar atoms with dominating dipolar interaction the   quasi-2D solitons  collapse due to an excess of dipolar energy as in a trapped dipolar BEC \cite{jbohn} 
and when the contact atomic repulsion dominates over the dipolar interaction the atoms escape to infinity and no soliton can be formed.
The result of the present study is illustrated using the numerical solution of the
time-dependent three-dimensional (3D) mean-field Gross-Pitaevskii (GP) equation \cite{ska} using realistic values of the interaction 
parameters  of $^{164}$Dy atoms \cite{ExpDy}.  

These quasi-2D dark-in-bright solitons are themselves bright solitons mobile in the $x-z$ plane
and are excited states of the quasi-2D bright solitons  with a notch (zero in density), for example,  along $x=0$ or $z=0$. The  dark-in-bright solitons  with a zero in the middle
are similar to the lowest odd-parity excited states of the harmonic oscillator potential. 
 Dark solitons in a confined BEC have been observed experimentally \cite{dark,dark1,dark3}.
The dark solitons with a notch imprinted \cite{dark,dark2}
on a trapped  repulsive BEC are dynamically unstable and are destroyed by snake instability  \cite{stabi3,stabi4,stabi5,stabi11,stabi12,stabi13,inst2,inst3}.  Such instability can be reduced as  the confining trap is made weaker \cite{inst2}. The present  dark-in-bright solitons created on  quasi-2D solitons are not subject to any confining trap in the $x-z$ plane and do not exhibit 
any kind of dynamical instability.

To study the stability of the quasi-2D solitons we performed two tests. 
The solitons are found to exhibit sustained breathing oscillation upon a
small perturbation, introduced by a change in the scattering length, confirming their stability. Such a change in scattering length can be realized experimentally 
by varying a uniform background magnetic field near a Feshbach resonance \cite{fesh}. 
As a more stringent test, we demonstrate the  generation of a dark-in-bright 
soliton   by introduzing an extra phase of $\pi$ on one half of the wave 
function (phase imprinting). Such a phase can be introduced in a laboratory   using 
the dipole potential of a far detuned laser beam \cite{dark,dark2}. In a numerical real-time simulation 
of the mean-field model such a phase-imprinted profile is used as the initial state,
which quickly transforms into a quasi-2D dark-in-bright soliton without 
exhibiting any dynamical instability at large times.

At  medium velocities the frontal 
collision between two quasi-2D solitons is found to be quasi elastic
while the two solitons come out without any visible deformation in shape. 
Only in 1D the collision between two analytic solitons is truly elastic. However, 
inelastic nature of the collision is expected at very low velocities in the case of
two quasi-2D dipolar solitons.

%{ In Sec. II the time-dependent 3D mean-field model
%for the dipolar BEC soliton is presented.
% The
%results of numerical calculation are exhibited in Sec. III.
%     Finally, in Sec. IV a
%brief summary of our findings is presented.
%}

\section{Mean-field model}  
   \label{II}

At ultra-low temperatures the properties of a dipolar condensate of $N$ 
atoms, each of mass $m$, can be described by the mean-field 
GP equation, for the wave function $\phi({\mathbf r},t)$,
with nonlocal nonlinearity of the 
form:~\cite{cr,Santos01}
\begin{eqnarray}
i\hbar\frac{\partial \phi({\mathbf r},t)}{\partial t} &&= 
\biggr[-\frac{\hbar^2}{2m}\nabla^2+V_{\mathrm{trap}}({\mathbf r}) + \frac{4\pi\hbar^2a N}{m}| 
\phi({\mathbf r},t)|^2 \nonumber \\&& + N \int U_{\mathrm{dd}}({\mathbf  r}-{\mathbf r}') 
\left\vert\phi({\mathbf r}',t)\right\vert^2 d{\mathbf r}' 
\biggr]\phi({\mathbf r},t),
\label{eqn:dgpe}
\end{eqnarray}
where $\int d{\bf r}  \vert  \phi({\mathbf r},t) \vert  ^2=1$ and $a$ 
is the scattering length. 
To study the 2D solitons mobile in the $x-z$ plane we consider the following harmonic 
(H) or OL potential in the $y$ direction
{\begin{eqnarray} \label{ht}
V_{\mathrm{trap}}^{\mathbf{H}}({\mathbf r}) &=& \frac{1}{2} m\omega^2 y^2,\\ 
                          V_{\mathrm{trap}}^{\mathbf{OL}}({\mathbf r})
&=& s E_R \sin^2 \left(\frac{2\pi}{\lambda} y  \right),
\label{olt} 
\end{eqnarray}
respectively,}
where   
$\omega$ is  the  frequency of the harmonic trap, $s$ is the strength of the OL trap 
in units of recoil energy $E_R=h^2/(2m\lambda^2)$ where $\lambda $ is the 
wave length of the lattice.  
The dipolar interaction, for magnetic dipoles, is \cite{cr} 
\begin{eqnarray}
U_{\mathrm{dd}}(\bf R)\equiv \frac{\mu_0 \mu^2}{4\pi}   V_{\mathrm{dd}}(\bf R) =\frac{\mu_0 \mu^2}{4\pi}\frac{1-3\cos^2 \theta}{ \vert  {\bf R} \vert  ^3},
\end{eqnarray}
where ${\bf R= r -r'}$ determines the relative position of dipoles and $\theta$ 
is the angle between ${\bf R}$ and the direction of polarization $z$, 
 $\mu_0$ is the permeability of free space 
and $ \mu$ is the dipole moment of an atom. The strength  of  dipolar interaction can be  expressed in terms  
 of a dipolar length  $a_{\mathrm {dd}}$ 
  defined  by  \cite{cr}
\begin{equation}
a_{\mathrm{dd}}\equiv\frac{ \mu_0  \mu^2  m}{12\pi \hbar^2}.
\end{equation}

For the formation of a  quasi-2D soliton, mobile  in the $x-z$ plane,   
a dimensionless GP equation for the dipolar BEC  soliton
can be written as   \cite{cr,1DA}
\begin{eqnarray}&& \,
 i \frac{\partial \phi({\bf r},t)}{\partial t}=
{\Big [}  -\frac{\nabla^2}{2 }+V(y)
+ g \vert \phi({\bf r},t) \vert^2
\nonumber \\  &&  \,
+ g_{\mathrm {dd}}
\int V_{\mathrm {dd}}({\mathbf R})\vert\phi({\mathbf r'},t)
\vert^2 d{\mathbf r}' 
{\Big ]}  \phi({\bf r},t),
\label{eq3}
\end{eqnarray}
with 
{
\begin{eqnarray}
V^{\mathbf{H}}(y) = \frac{1}{2}y^2,   \quad ‏V^{\mathbf{OL}}(y)= \frac{s}{2}\sin^2y,
\end{eqnarray}
for} the harmonic trap (\ref{ht}) and the 
OL trap (\ref{olt}), respectively,
where 
$g=4\pi a N,$ 
$g_{\mathrm {dd}}= 3N a_{\mathrm {dd}}$.  
In the case of the harmonic trap,  length is expressed in  (\ref{eq3}) in units of 
oscillator length  $l=\sqrt{\hbar/(m\omega)}$, 
energy in units of oscillator energy  $\hbar\omega$, probability density 
$|\phi|^2$ in units of $l^{-3}$, and time in units of $ 
t_0=1/\omega$. 
 In the case of the OL trap,  length is expressed  in units of 
  $l=\lambda/(2\pi)$, 
energy in units of recoil  energy  $E_R$, probability density 
$|\phi|^2$ in units of $l^{-3}$, and time in units of $ 
t_0=m\lambda^2/(2\pi h)$.

 \section{Numerical Results}
  We consider $^{164}$Dy
atoms in this study of  BEC solitons. 
  The magnetic moment of a $^{164}$Dy  
atom is  $ \mu = 10\mu_B$
\cite{ExpDy} 
with 
$\mu_B$ the Bohr magneton leading to the dipolar lengths $a_{\mathrm {dd}}(^{164}$Dy$) \approx 132.7a_0$,    with $a_0$ the Bohr radius. 
The dipolar interaction in $^{164}$Dy
atoms is roughly   eight times larger than that in $^{52}$Cr
atoms with a dipolar length  $a_{\mathrm {dd}}(^{52}$Cr)$\approx 15.3a_0$ 
\cite{cr}. Because of the large magnetic moment, 
the dipolar  $^{164}$Dy BEC will facilitate the formation of bright and 
dark-in-bright solitons.

% We take $l\equiv \sqrt{\hbar/m\omega} =1$ $\mu$m. In a  $^{164}$Dy BEC this corresponds to an angular trap frequency $\omega =  2\pi \times 61.6$ Hz corresponding to $t_0= 2.6$ ms. 

{  The time-dependent GP equation (\ref{eq3}) 
can be solved by different numerical methods \cite{CPC1,CPC3}.
We solve the GP equation  
by the split-step 
Crank-Nicolson method \cite{CPC1,CPC2}
with real- and imaginary-time propagation
  in  Cartesian spatial coordinates  
using a space step of 0.1  
and a time step of 0.0002    for a soliton of smaller size and a space step of 0.2  
and a time step of 0.001    for a soliton of larger size. We employ the computer 
programs of Refs. \cite{ska,ska2,ska3,CPC2} for this purpose. }
As the bright and dark-in-bright solitons are the ground states with specific parity, 
they can be obtained by imaginary-time propagation. The stability of the 
solitons is   studied by real-time propagation. 
% The dipolar potential term is treated by Fourier transformation  in momentum space as \cite{ska}
%\begin{eqnarray}
%\widetilde V_{\mathrm{dd}}({\bf k})= \frac{4\pi}{3}\left(\frac{3k_z^2}{{\bf k}^2}-1   \right)
%\end{eqnarray}
%and the nonlocal dipolar integral in the GP equation (\ref{eq3}) is treated by the 
%convolution 
% \cite{Santos01,ska}
%\begin{eqnarray}
%\int   V_{\mathrm{dd}}({\bf r-r'})n({\bf r'}) = \int \frac{d\bf k}{(2\pi)^3}e^{-i{\bf k}\cdot {\bf r}} \widetilde V_{\mathrm{dd}}({\bf k})\widetilde n({\bf k}), 
%\end{eqnarray}
%where $n({\bf r})
%=|\phi({\bf r})|^2$ and tilde represents Fourier transformed quantities. 

To obtain a bright soliton  in numerical simulation  we consider the following initial
Gaussian   wave function
\begin{equation}\label{bright}
\phi({\bf r})= \frac{\pi^{-3/4}}{\sqrt{w_x w_y w_z}}\exp \Big[-\frac{x^2}{2w_x^2}  -\frac{y^2}{2w_y^2}-\frac{z^2}{2w_z^2}  \Big],
\end{equation}
where $w_x, w_y$ and $w_z$ are the widths along $x$, $y$ and $z$ axes, respectively. For a
dark-in-bright soliton, for example,  with a notch along $x=0$, we consider the initial function 
\begin{equation}\label{dark}
\phi({\bf r})= \frac{\sqrt 2 x}{\sqrt{\pi^{3/2} w_x^3 w_y w_z  }}\exp \Big[-\frac{x^2}{2w_x^2}  -\frac{y^2}{2w_y^2}-\frac{z^2}{2w_z^2}  \Big].
\end{equation}

 \begin{figure}[!t] 

\begin{center}
\includegraphics[width=.49\linewidth]{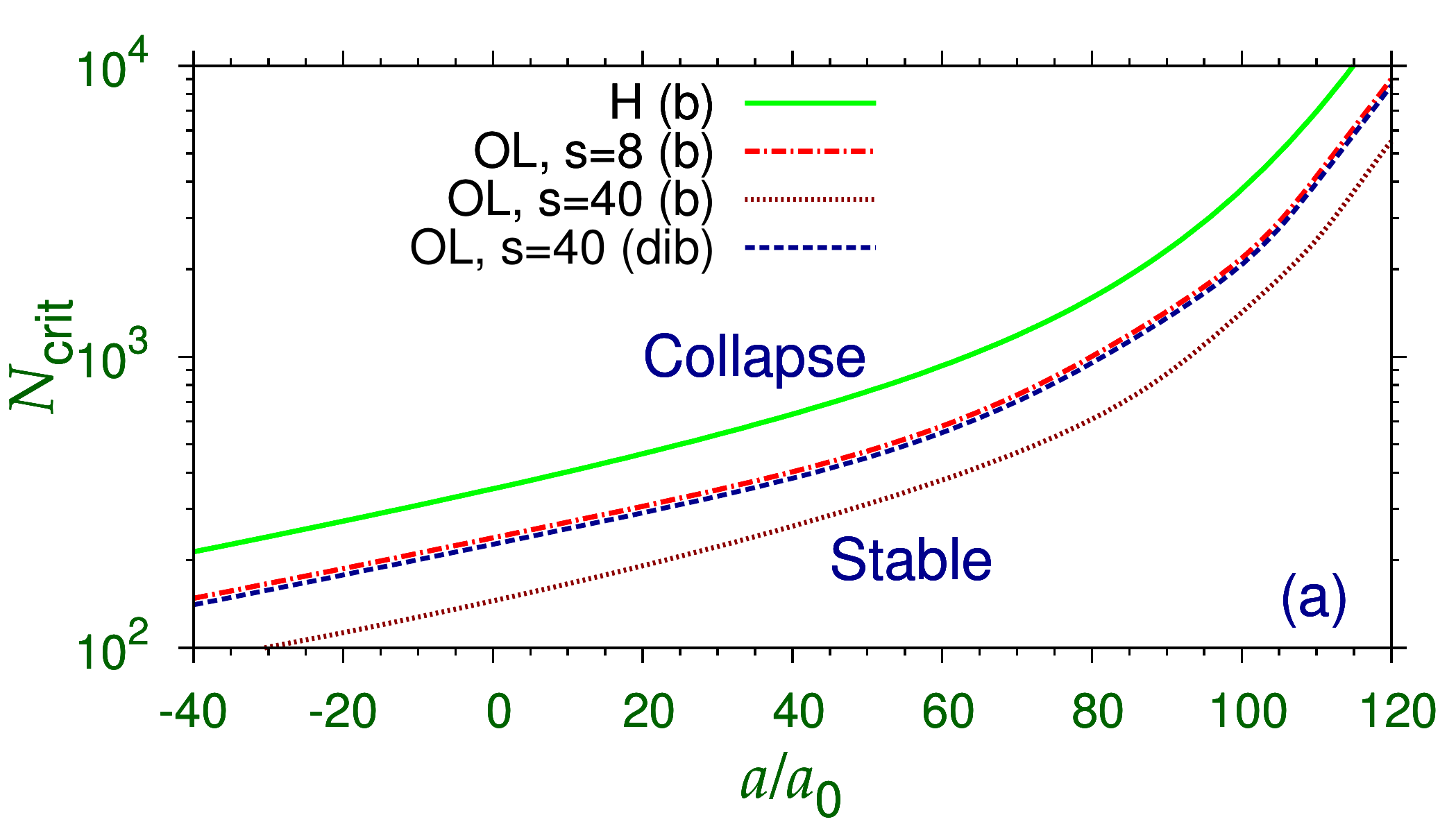} 
 \includegraphics[width=.49\linewidth]{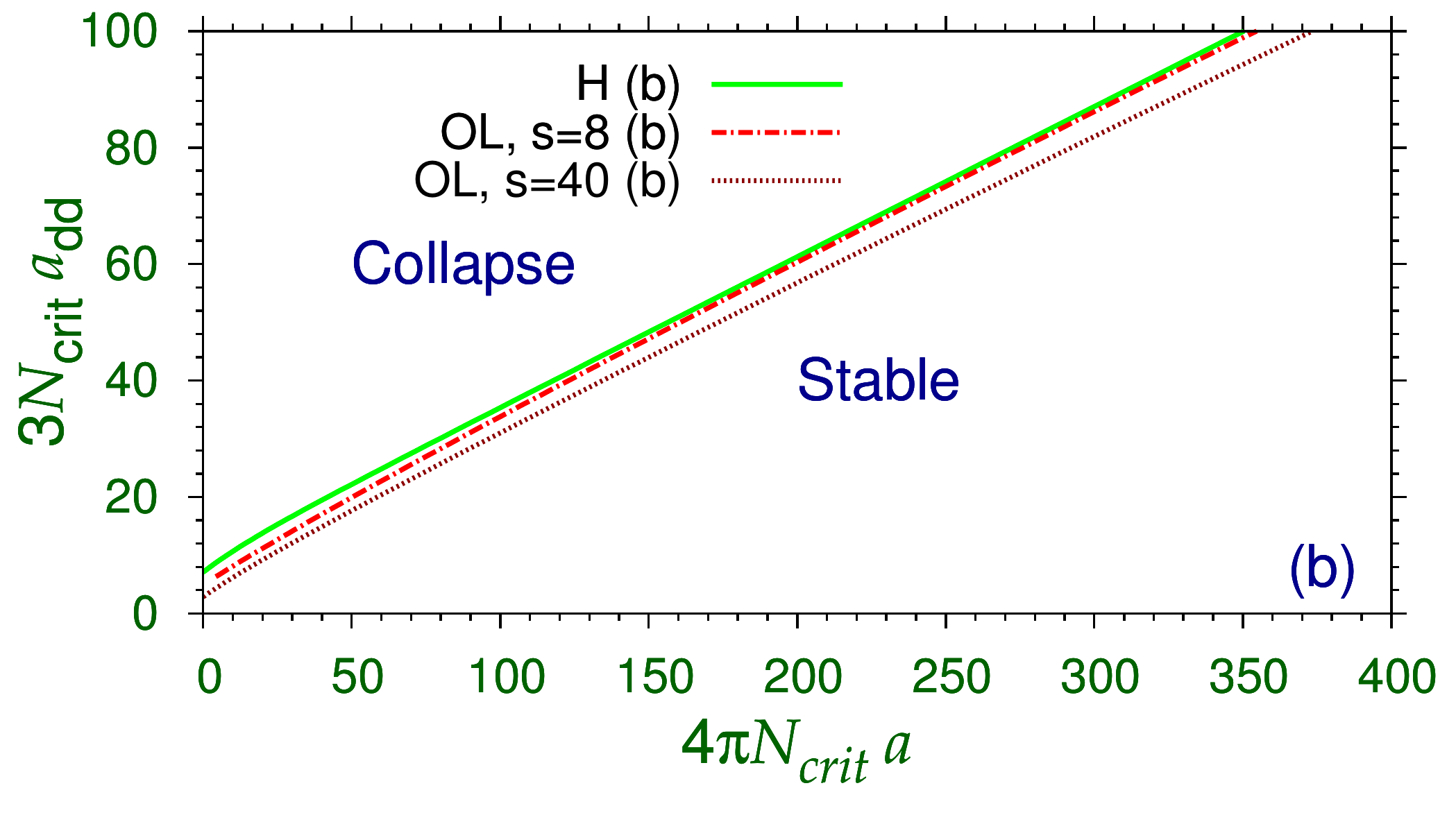} 

\caption{ (Color online)  (a) A $N_{\mathrm{crit}}-a$ phase plot illustrating the parameter space for the 
formation of a stable quasi-2D bright (b) and dark-in-bright (dib) 
soliton in a $^{164}$Dy BEC with $a_{\mathrm{dd}}=132.7a_0$ for harmonic (H) and OL traps.
The soliton is stable for $N<N_{\mathrm{crit}}$ and collapses for $N>N_{\mathrm{crit}}$.
(b)  A universal $g_{\mathrm{dd}}-g$ phase plot  showing the stability region of a quasi-2D bright soliton
applicable 
to a BEC of any dipolar atom. All quantities plotted in this and other figures are 
dimensionless. In all calculations of this paper
 the length scale $l=1$ $\mu$m.
%  Elastic collision dynamics of two solitons of figure \ref{fig2} (c) from a solution of the  
%3D model (\ref{eq3}) of two identical solitons of figure \ref{fig2} (c) in opposite directions
%via a plot of linear density $n_{1D}(x,t)$ versus $x$ and $t$.
}
\label{fig1}

 \end{center}

\end{figure}

A quasi-2D bright soliton was obtained by solving  
the 3D GP equation (\ref{eq3}) by imaginary-time propagation using 
the initial function (\ref{bright}) with conveniently chosen widths
for different values of  number of atoms $N$ and 
scattering length $a$. For a dominating dipolar interaction ($a_{\mathrm{dd}}>a$), 
the bright solitons are possible for a moderate number of atoms. The system collapses as
the number of atoms is increased beyond a critical number $N_{\mathrm{crit}}$
due to an excess of dipolar energy density \cite{jbohn}.   
For a dominating contact repulsion  ($a_{\mathrm{dd}}<a$), no solitons can be formed. 
The 
stability region for the formation of a quasi-2D soliton in the parameter space for $^{164}$Dy atoms { is obtained from a imaginary-time propagation of  (\ref{eq3}) 
in the same way as  in Refs. \cite{g1,g2} for a nondipolar trapped BEC. For a small  number of atoms ($N\sim 100$) and 
scattering length $a$ ($<a_{\mathrm{dd}}$) a quasi-2D bright soliton is numerically 
obtained by  imaginary-time propagation starting from the  initial Gaussian profile (\ref{bright}). 
Then the   calculation is repeated increasing the number of atoms $N$ keeping all 
other parameters fixed. No such soliton can be obtained numerically for $N$ greater than the critical  number $N_{\mathrm{crit}}$.
The result of our finding
is illustrated  in the $N_{\mathrm{crit}}-a$ phase plot of figure \ref{fig1}(a)
showing the critical number $N_{\mathrm{crit}}$
 for a harmonic trap as well as an OL trap with two 
strengths: $s=8$ and 40. 
We also plot in this figure 
the critical number of the dark-in-bright soliton on the OL potential of strength 
$s=40$ obtained by imaginary-time simulation starting from the initial function (\ref{dark}) in a similar fashion.}
Although, the case of the harmonic potential has been studied before and in the following 
we present  only results for the OL potential, we include the results of the harmonic 
potential in the phase plot of figure \ref{fig1}  to address the difference between the two cases.  
The critical number of atoms 
increases with the increase of scattering length $a$.  
The bright solitons are unconditionally stable and 
last for ever in real-time propagation without any visible change of shape.  
Although, the stability plot of figure \ref{fig1}(a) is closely related to the parameters of  $^{164}$Dy atoms, a more universal  $g-g_{\mathrm{dd}}$ plot 
as shown in figure  \ref{fig1}(b) is applicable to any dipolar BEC. The universal stability 
line of  figure  \ref{fig1}(b) has the approximate  straight line shape and can be easily extrapolated to the case of  larger nonlinearities $g$ and $g_{\mathrm{dd}}$.

{
The  1D harmonic and OL potentials along $y$ direction are quite different and it is of interest to know how this changes the stability and shape of the quasi-2D solitons. We 
will mostly study the quasi-2D OL solitons in a deep OL trap when the quasi-2D soliton 
occupies a single OL site. The harmonic trap is wide  and can accommodate a larger number of atoms in a larger spatial extention compared to the OL trap which can accommodate a smaller number of atoms in a single site with a smaller spatial extension. For a large number of atoms in the same OL trap the dipolar energy density will be too large to provoke 
collapse instability \cite{jbohn}, viz. figure \ref{fig1}. However, as the strength $s$ of the OL is lowered from 40 to 8, some atoms can tunnel to the adjescent site. This reduces the central   
dipolar energy density and collapse instability, thus accommodating a larger number of atoms
in a shallow OL potential  as can be seen in figure  \ref{fig1}. The dark-in-bright soliton 
is an excited state of the bright soliton with a zero in the middle and has a much larger spatial extension with a smaller dipolar energy density and hence can accommodate a larger 
number of atoms as can be seen by comparing the plots marked ``b" and ``dib" for $s=40$
in figure   \ref{fig1}(a). The shape (aspect ratio in the 2D plane and not the actual size)
of the quasi-2D solitons 
in the case of a harmonic and deep OL potential 
is determined completely by the dipolar interaction and not the confining trap.  
However, the actual size of these solitons will be determined by both the confining potential and dipolar interaction.
Hence these quasi-2D solitons have a similar shape (not size) in the 2D plane, independent of the trapping potential. For a shallow OL potential
quasi-2D solitons occupying a few OL sites can be formed close to the stability line  in figure \ref{fig1}. 
To the right and away from the stability line,  the atomic contact repulsion increases and the soliton may 
extend to several OL sites and eventually will be delocalized for an excess of contact repulsion. The quasi-2D solitons on a harmonic potential has been thoroughly studied 
before \cite{2D1,2D2,2D3,2D4,2dcol} and we present a comprehensive study of the quasi-2D solitons on an OL potential in the following.
}

 \begin{figure}[!t]

\begin{center}
\includegraphics[width=.32\linewidth,clip]{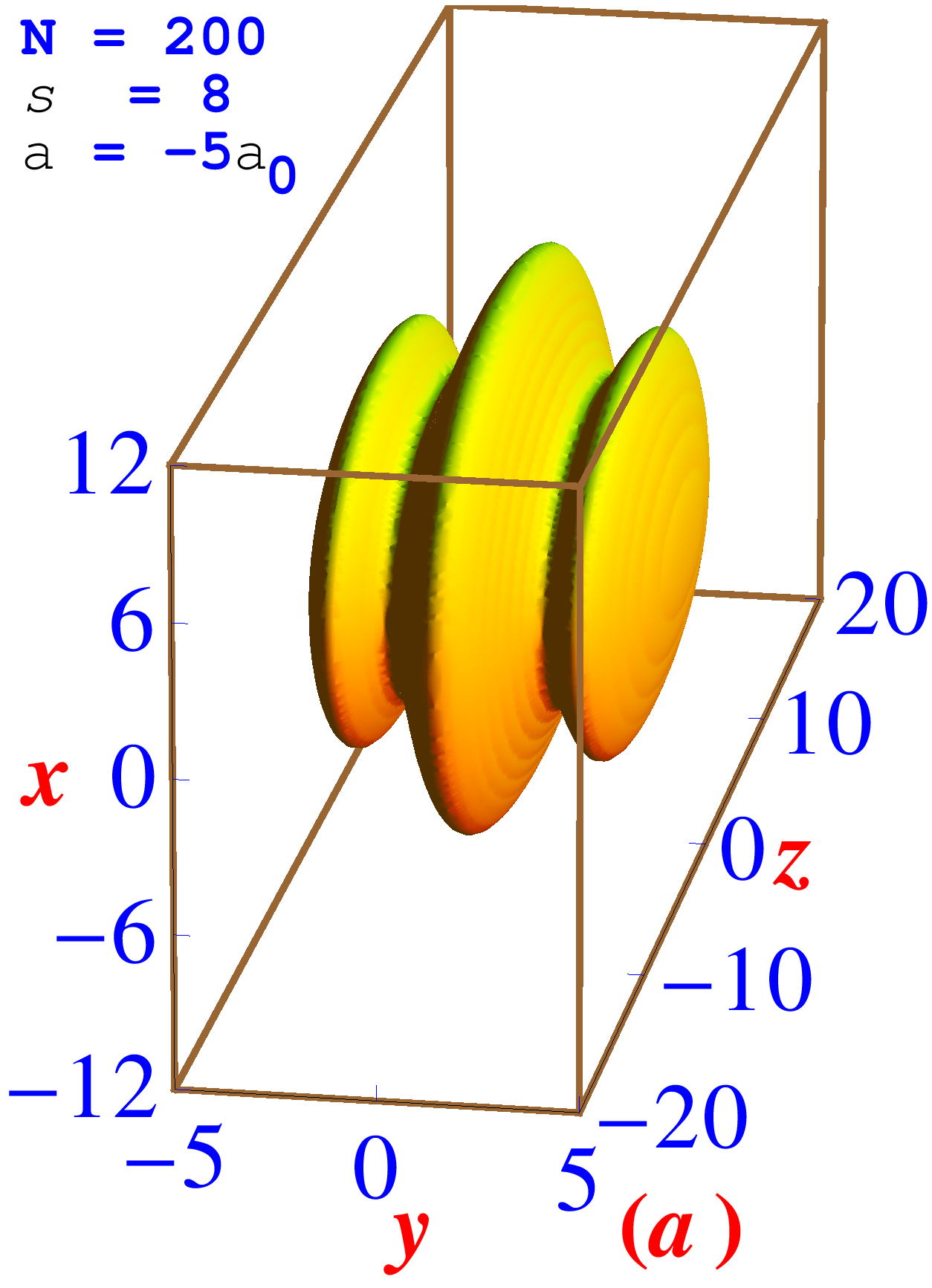}
\includegraphics[width=.32\linewidth,clip]{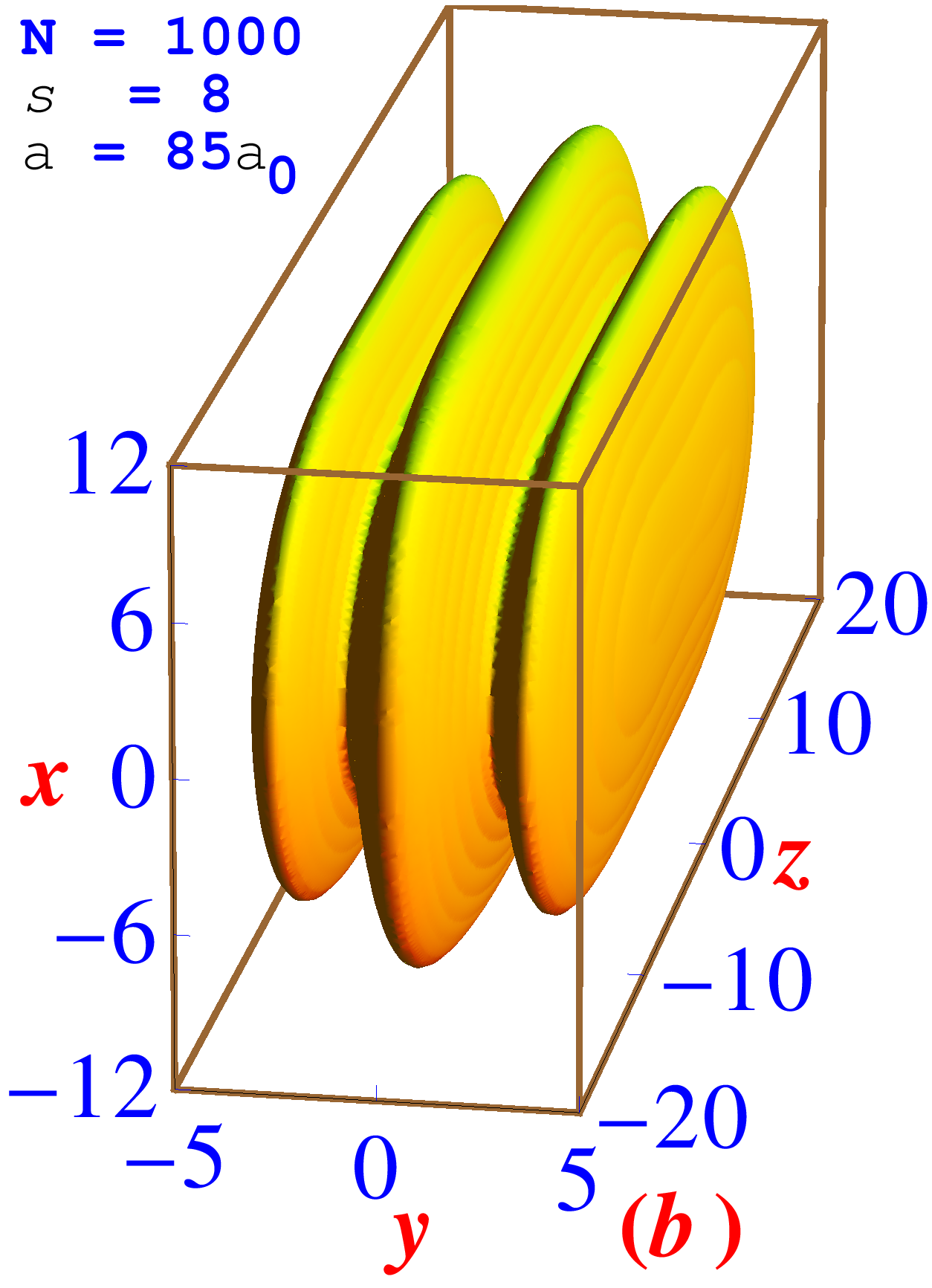}
\includegraphics[width=.32\linewidth,clip]{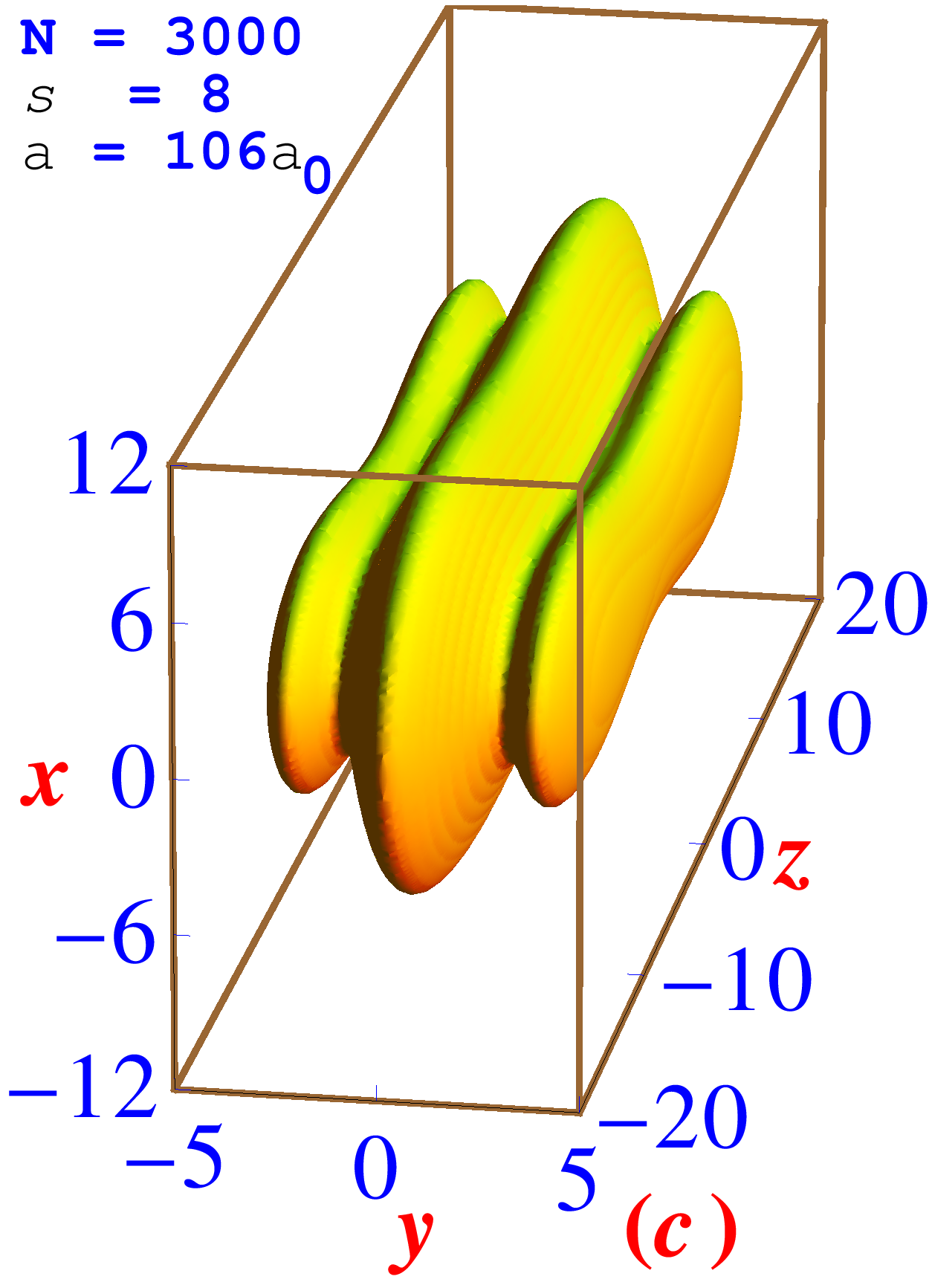}
\includegraphics[width=.32\linewidth,clip]{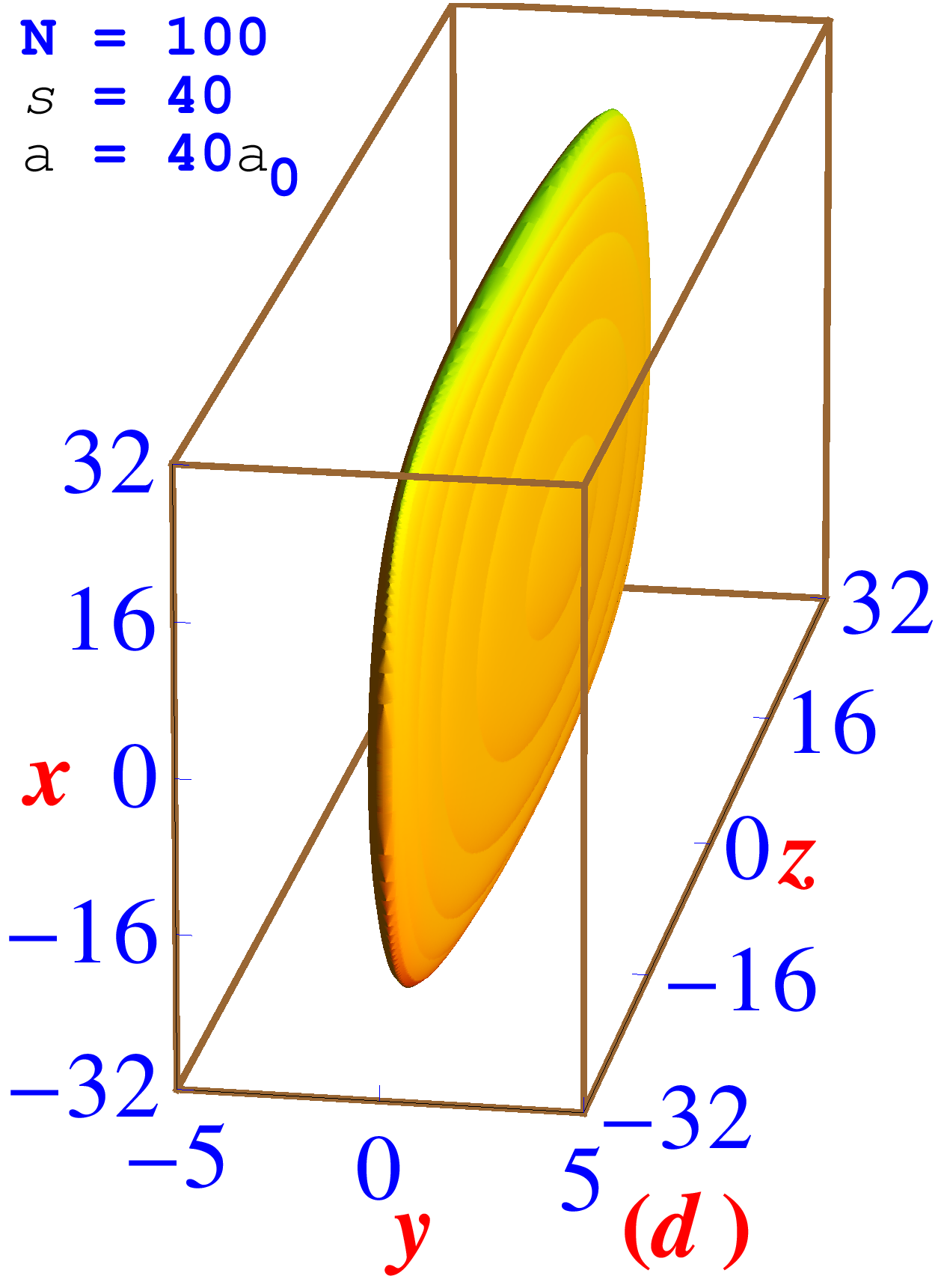}
\includegraphics[width=.32\linewidth,clip]{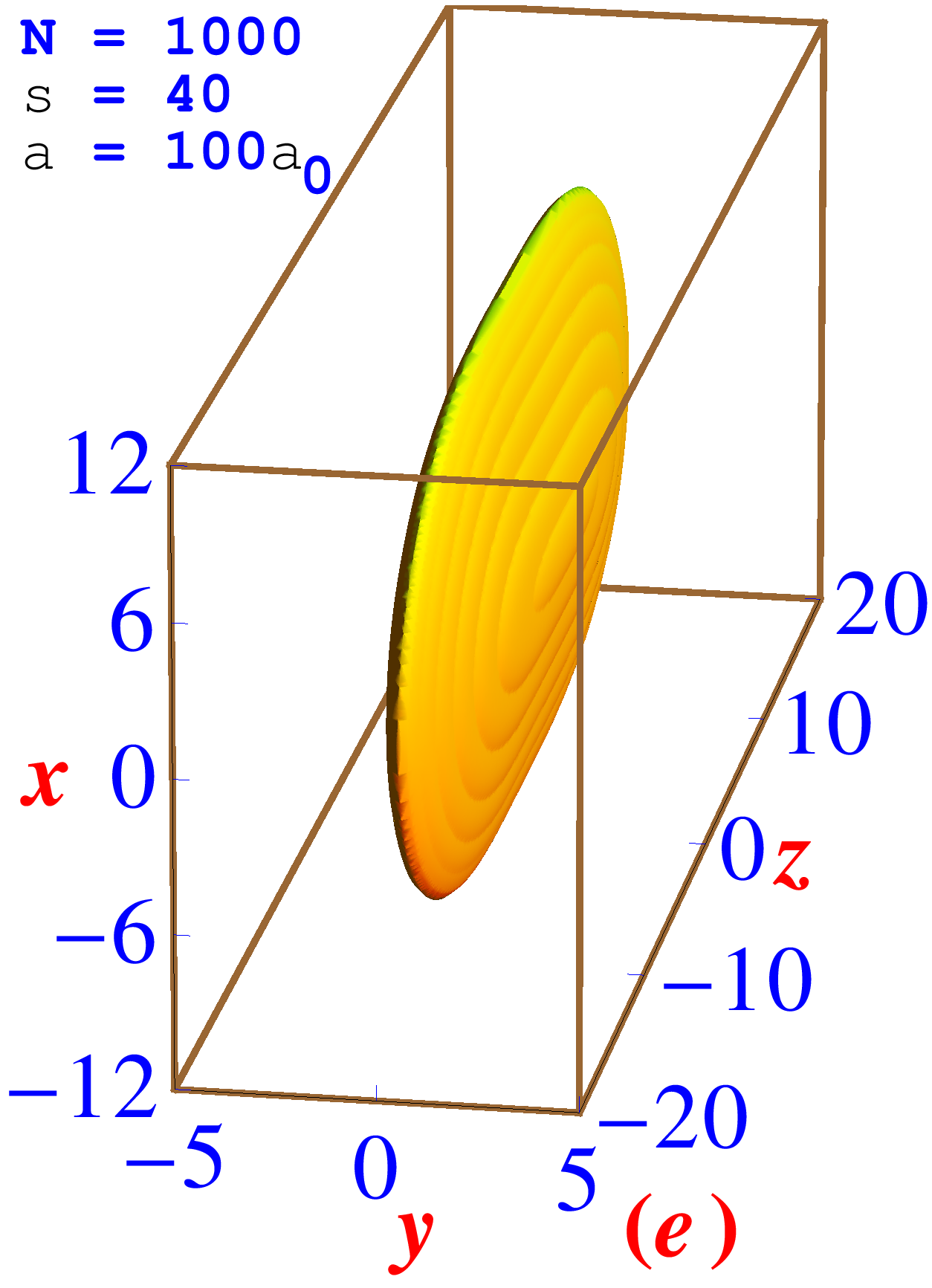}
\includegraphics[width=.32\linewidth,clip]{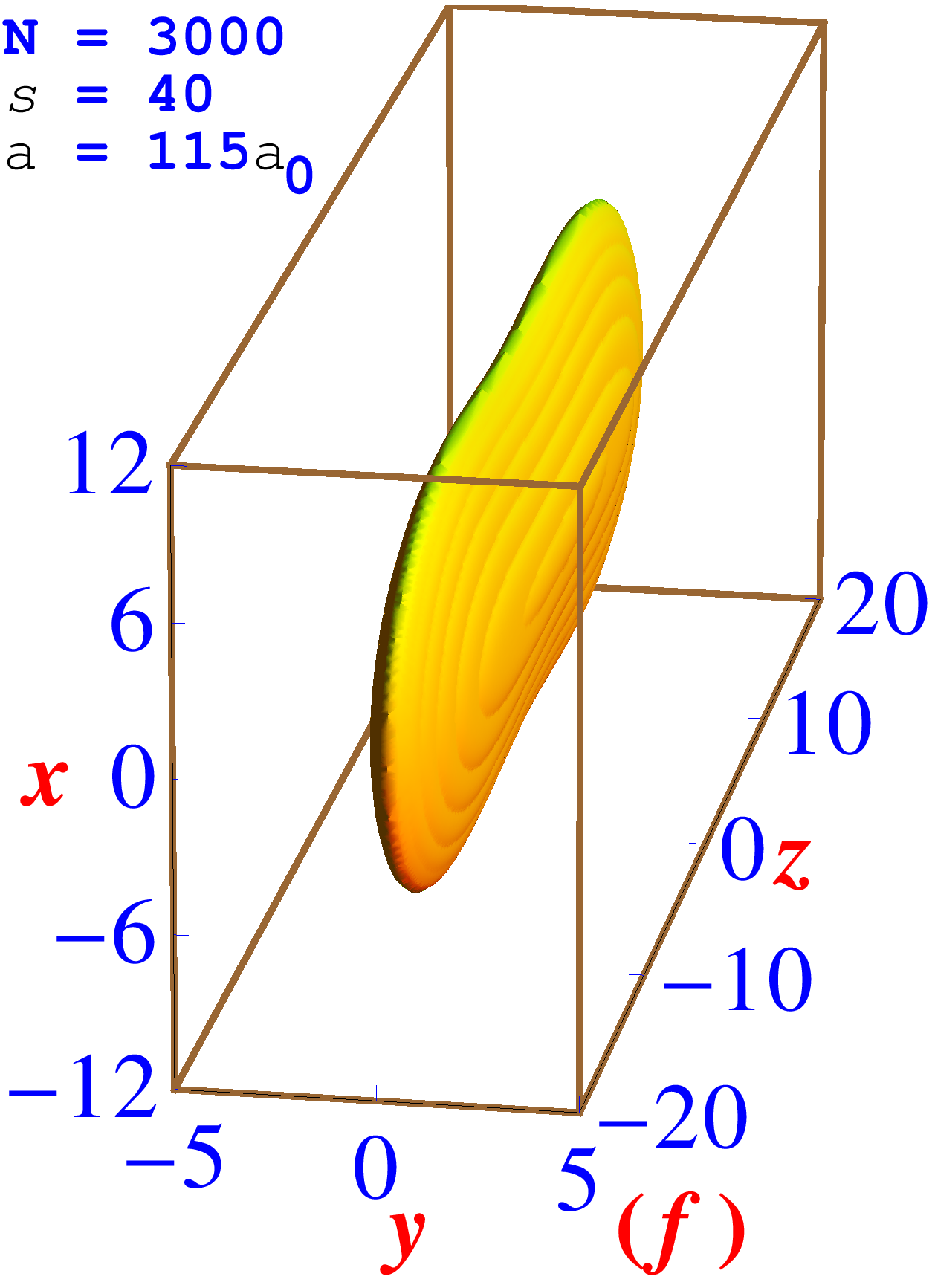}

\caption{ (Color online) 3D isodensity contour   $|\phi({\bf r})|^2$ of a bright soliton of  
$^{164}$Dy atoms in the OL trap $V(y)=  s\sin^2y/2$
for (a) $a= -5a_0, N =200, s =8$,  (b)  $a= 85a_0, N= 1000, s=8$,  (c)  $a= 106a_0, N= 3000, s =8$,  (d)  $a= 40a_0, N =100, s=40$,  (e)  $a= 100a_0, N =1000, s=40$, and  (f)  $a= 115a_0, N =3000, s =40$.
  The dimensionless
density on 
the contour is 0.000001. With present length scale $l=1$ $\mu$m 
this density  corresponds to  $ 10^{6}$ atoms/cm$^3$. The central density of the solitons is typically $ 10^{10}$ atoms/cm$^3$. }

\label{fig2} \end{center}

\end{figure}

In figures \ref{fig2} (a) $-$ (f) we display the 3D isodensity profile of the quasi-2D 
bright solitons for %$N=250, 500, 1000, $ and $2000$ and $a= 26a_0, 52a_0, 77a_0$, and $94a_0$, respectively. 
{differnet $N, a$ and $s$}, obtained by imaginary-time propagation 
of the GP equation with the OL potential
starting   with the initial wave function (\ref{bright}). 
 Because of the OL trap in the $y$ direction, the  
solitons have a quasi-2D shape in the $x-z$ plane, as can be seen in 
figure \ref{fig2}.  In this figure we present results for two strengths of the OL 
potential: $s=8$ and 40. Of these, $s=8$ corresponds to a weaker OL trap and the quasi-2D 
solitons extend over several sites of the OL potential. Compact quasi-2D solitons occupying 
a single site of the OL trap is possible for $s=40$. The size of the soliton in the 
$x-z$ plane  is small for parameters near the stability line, viz. figures \ref{fig2}(e) and (f), 
and is large for parameters away from it, viz. figure \ref{fig2}(d).

\begin{figure}[!t]

\begin{center}
\includegraphics[width=.24\linewidth,clip]{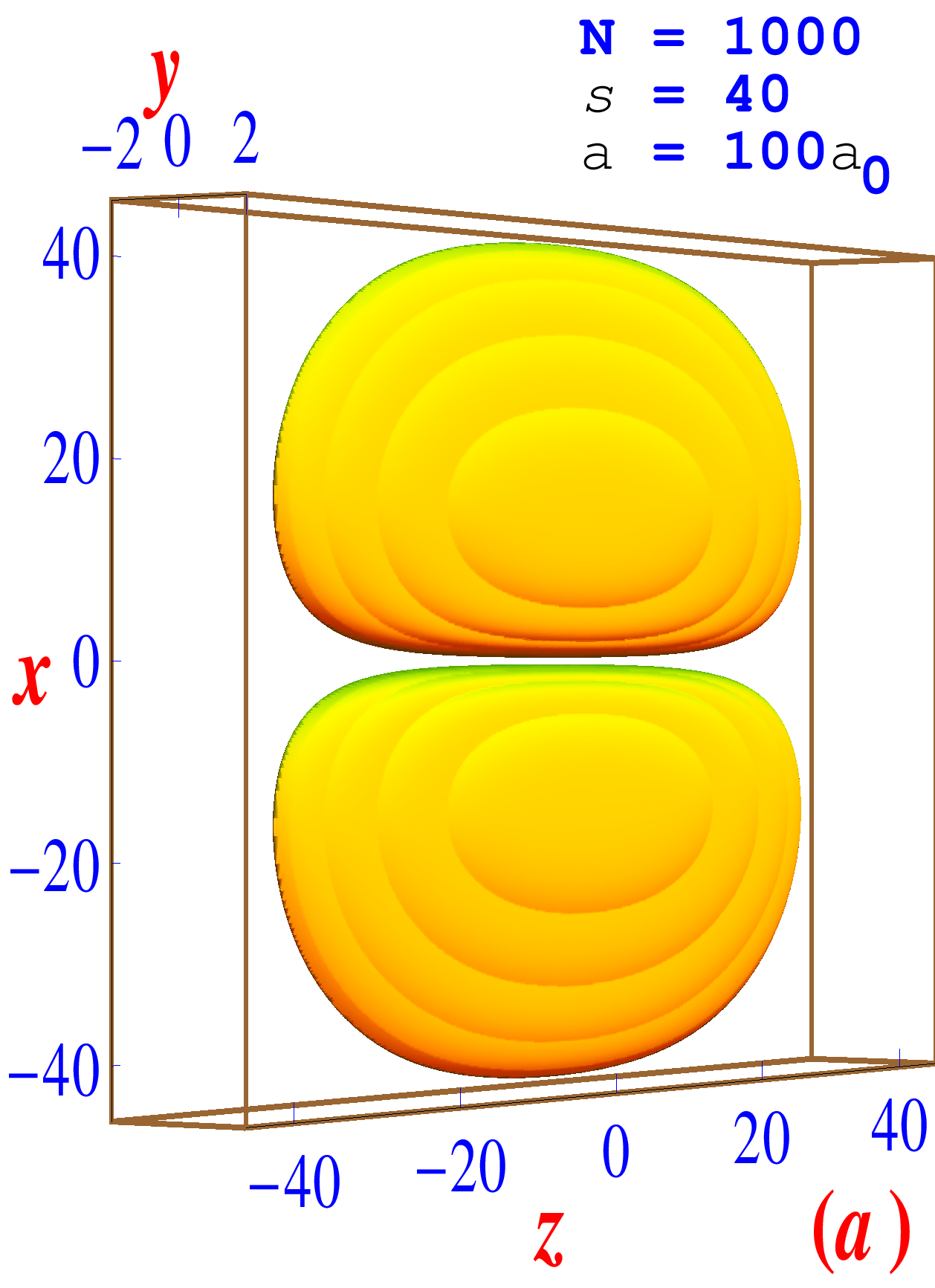}
\includegraphics[width=.24\linewidth,clip]{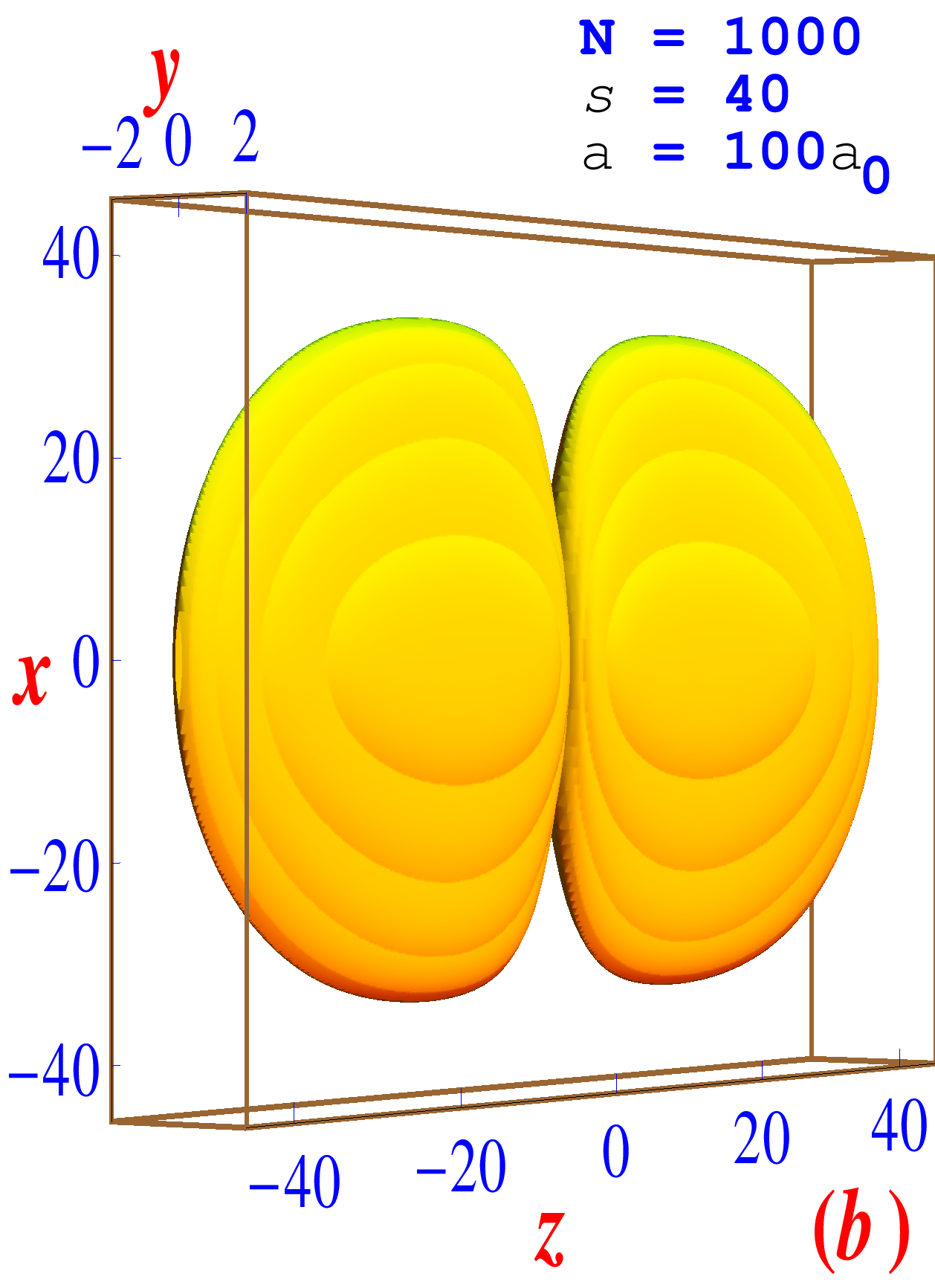} 
\includegraphics[width=.24\linewidth,clip]{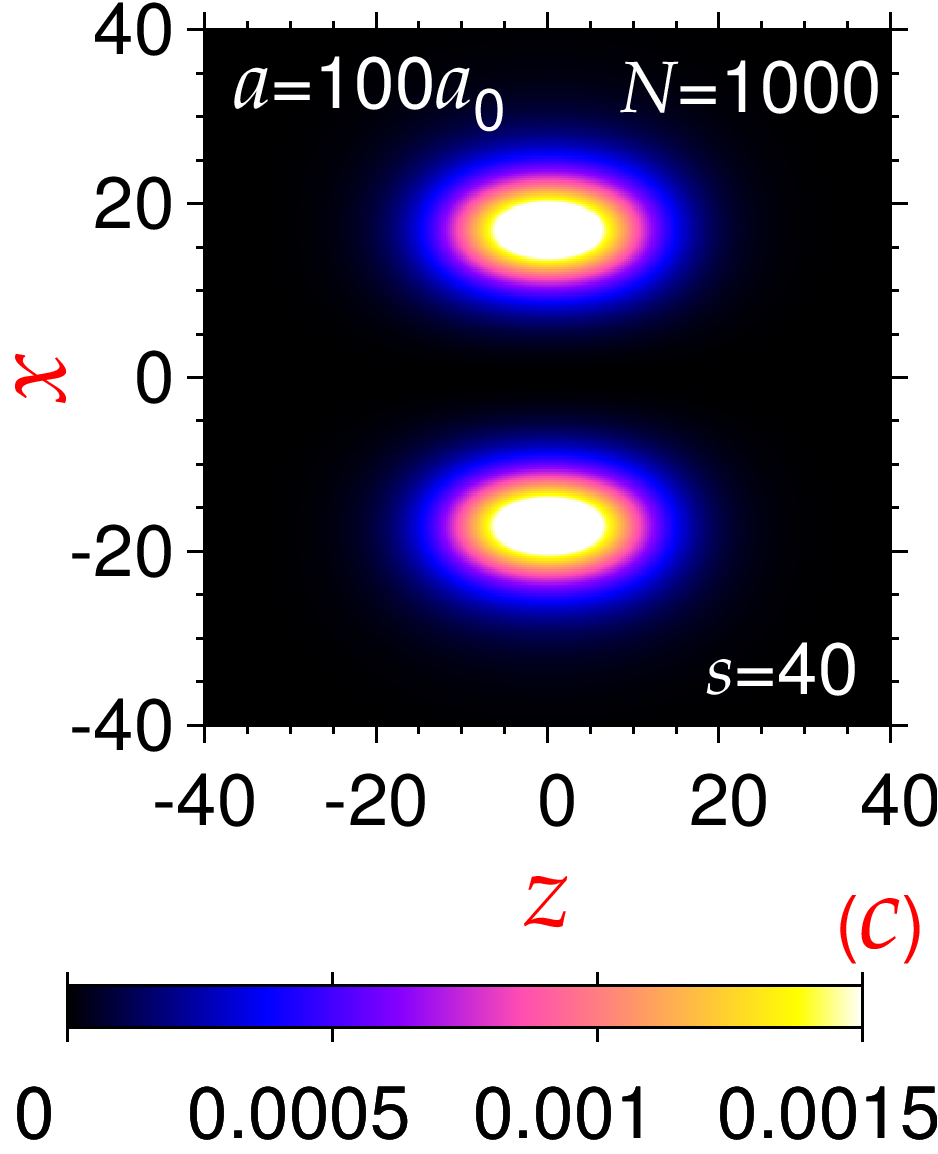}
\includegraphics[width=.24\linewidth,clip]{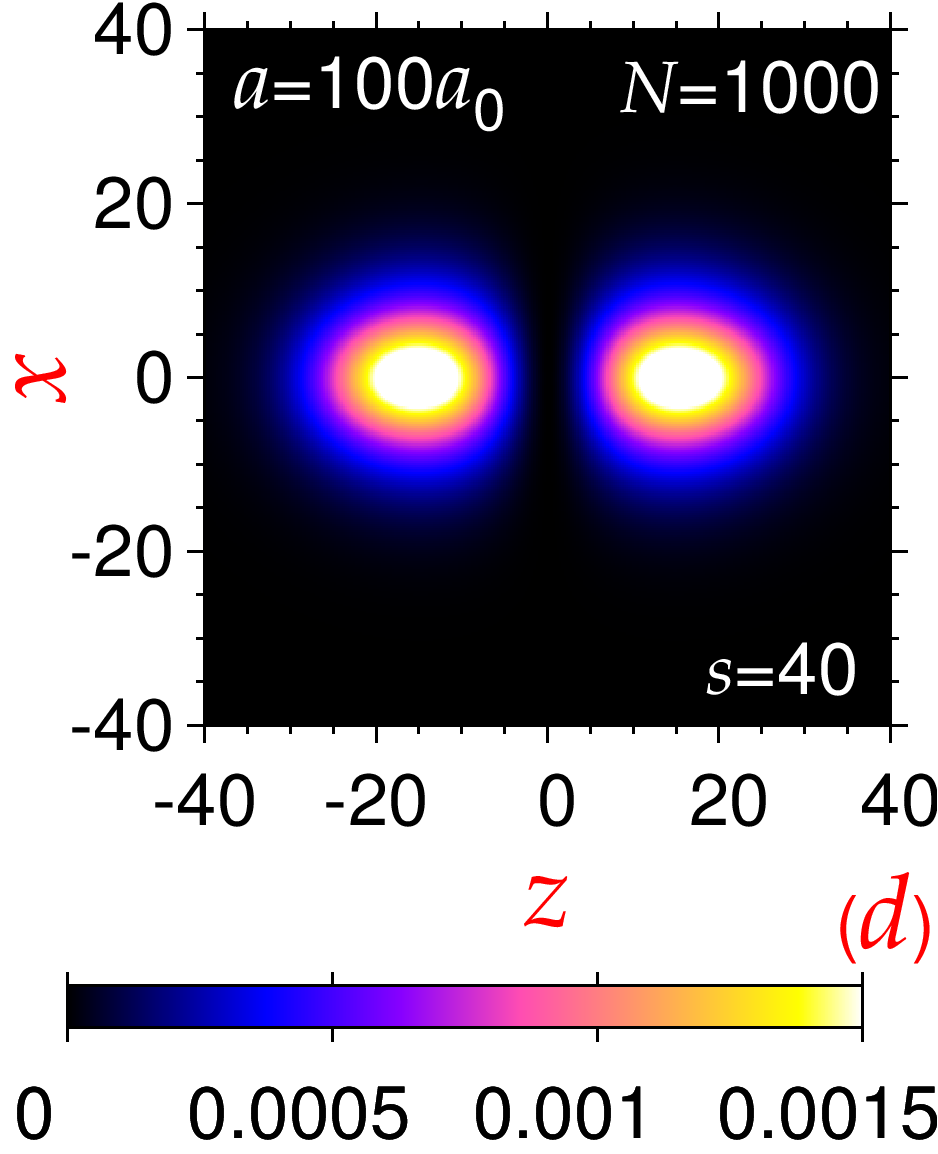} 
\caption{ (Color online) 3D isodensity contour   $|\phi({\bf r})|^2$ of a dark-in-bright soliton of  
$^{164}$Dy atoms in the OL trap $V(y)= s\sin^2y/2$ with $a= 100a_0, N =1000, s =40$, with a notch along 
(a) $x=0$ and (b) $z=0$. The density on the contour is 0.000001.
The contour plot of the 2D density $n_{2D}(x,z)$ of the two dark-in-bright solitons shown in (a) and (b),  respectively, are plotted in (c) and (d). 
}\label{fig3}
\end{center}

\end{figure}

{The present dark-in-bright solitons are the stable excited states of bright dipolar solitons in the same sense as the usual dark solitons of a trapped nondipolar BEC are very unstable  excitations of trapped BECs. There have been a large number of investigations about how to stabilize these nondipolar dark solitons \cite{stabi3,stabi4,stabi5,stabi11,stabi12,stabi13,inst2,inst3}. These studies revealed beyond any doubt that transverse snake instability of  
nondipolar dark solitons is an inherent property and cannot be eliminated. This casts doubt on the possibility of a decent experiment with these dark solitons. We demonstrate that the presence of dipolar interaction does not only make the dark solitons stable but also make them mobile in a plane. This opens a new scenario of performing precise experiments with the present dark-in-bright solitons. We will demonstrate how a stable dark-in-bright dipolar soliton can be realized in a laboratory by phase imprinting \cite{dark,dark2}
a bright soliton.}

In figures \ref{fig3} (a) $-$ (b) we exhibit the 3D isodensity contour of the quasi-2D 
dark-in-bright solitons for 
$N=1000,s=40$ and $a=100a_0$ with a notch along $x=0$ and $z=0$, respectively,
 obtained by a numerical solution
of the GP equation with the OL potential
starting   with the initial wave function (\ref{dark}). The notch in the density along 
the $x$ and $y$ axes is clearly visible in figures \ref{fig3} (a) and (b), respectively.
For the same set of parameters ($N,s$ and $a$), the dark-in-bright solitons
extend over a larger domain in the $x-z$ plane compared to the bright soliton, viz. figures \ref{fig2}(e) and  \ref{fig3}. In figures \ref{fig3}(c) and (d)  we show the contour plot  
of the 2D density 
\begin{eqnarray}
n_{2D}(x,z)=\int_{-\infty}^{\infty} dy |\phi({\bf r})|^2
\end{eqnarray}
of the two dark-in-bright solitons of figures \ref{fig3}(a) and (b), respectively. These 2D 
contour plots show clearly the density maxima of the solitons.

\begin{figure}[!t]

\begin{center} 
\includegraphics[width=.49\linewidth]{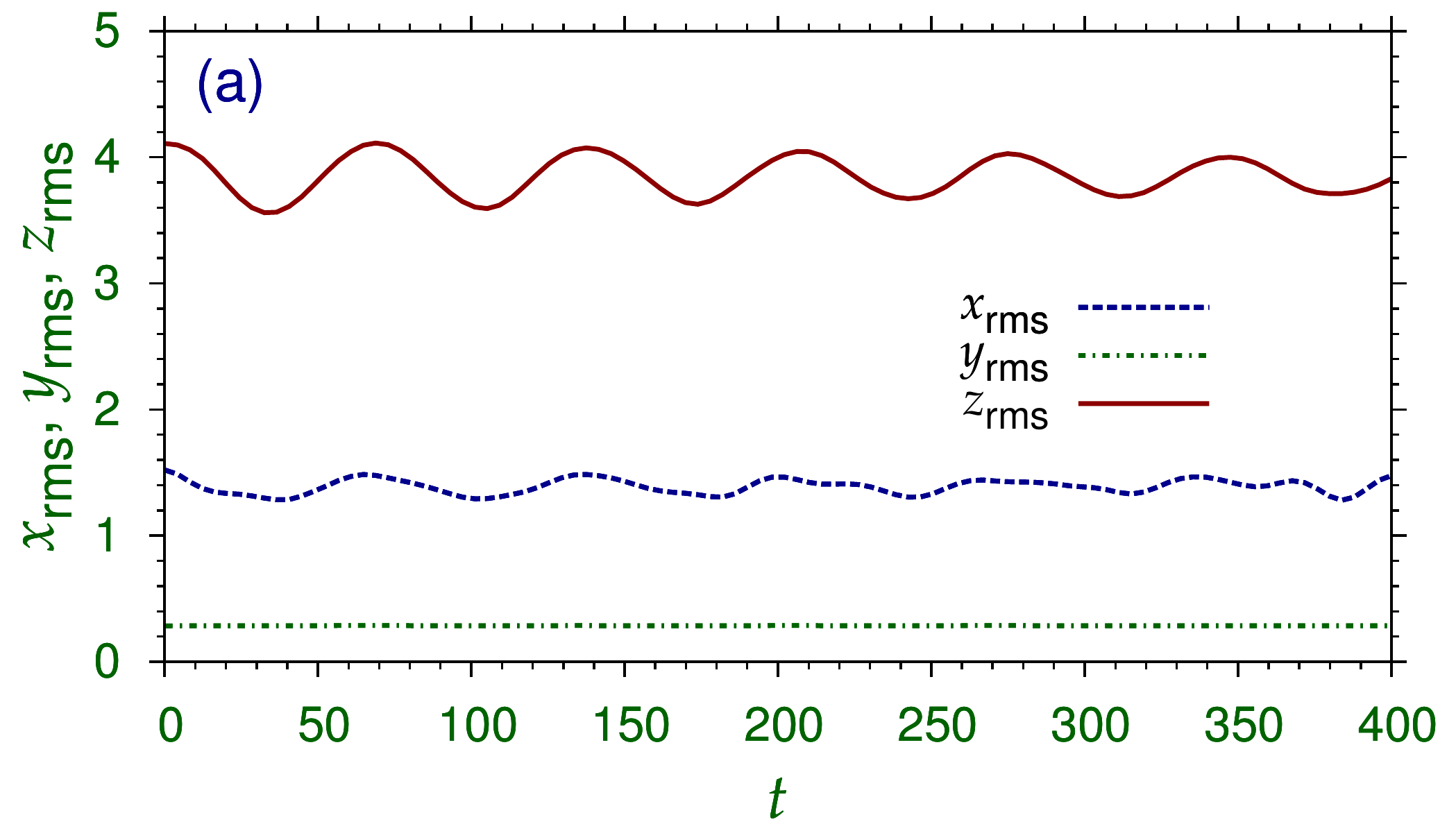}
 \includegraphics[width=.49\linewidth]{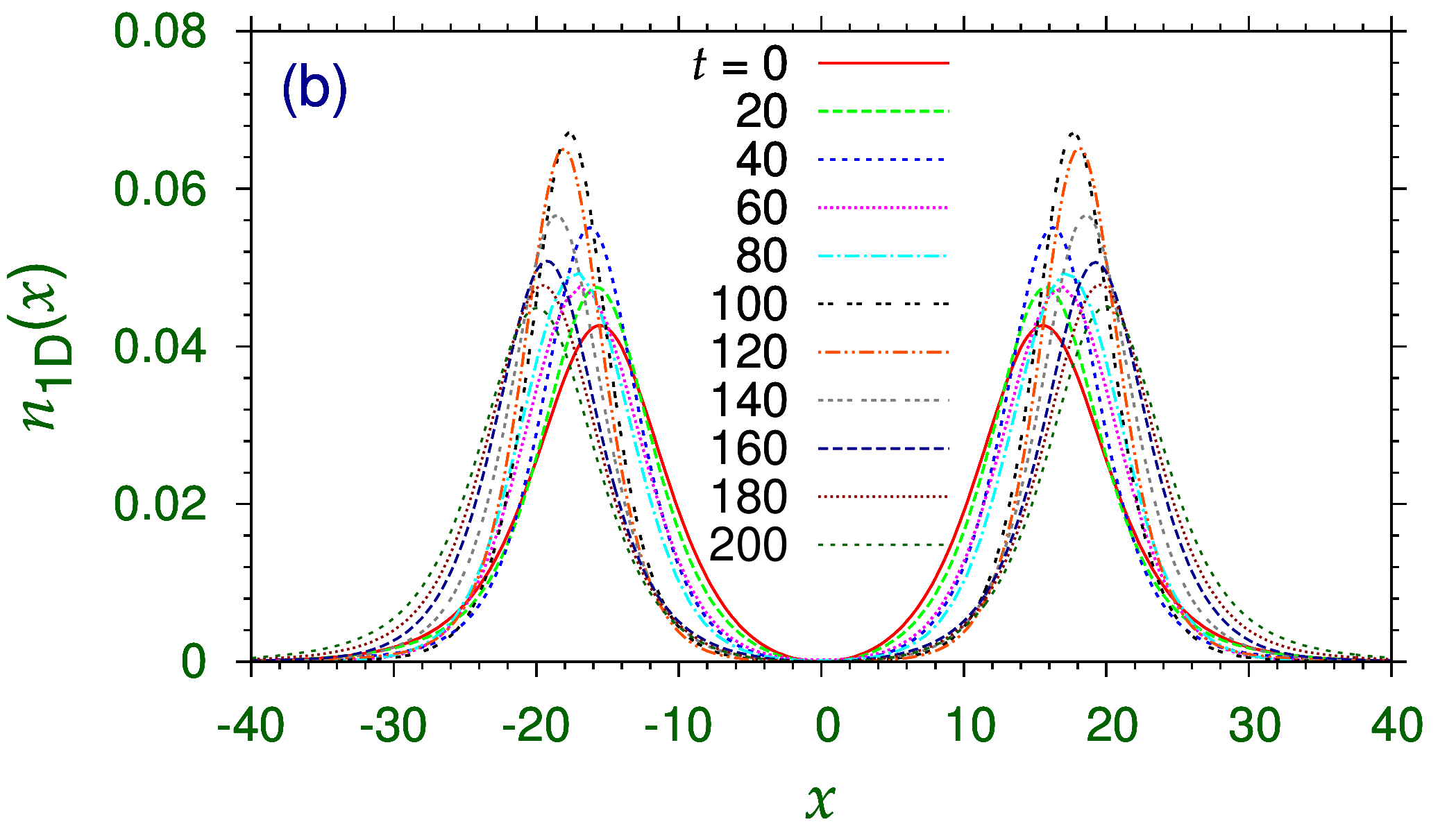}

\caption{ (Color online)   Stability test of a quasi-2D (a) bright and (b) dark-in-bright soliton with $N=1000, a=100a_0, s=40$. { (a) Plot of rms sizes $ x_{\mathrm{rms}} , y_{\mathrm{rms}} ,z _{\mathrm{rms}}$} vs. time $t$ of the bright soliton of figure \ref{fig2}(e)  during real-time evolution  of the bright soliton with the pre-calculated initial stationary state. 
 At the start of real-time evolution 
 at $t=0$ the scattering length $a$ is changed from $100a_0$ to $99a_0$.
(b)  Integrated 1D density $n_{1D}(x)$ vs. $x$  of the   dark-in-bright soliton of figure \ref{fig3}(a)
at different times  during real-time evolution  of the dark-in-bright soliton.   At $t=0$ the scattering length $a$ is changed from $100a_0$ to $95a_0$.
}\label{fig4} \end{center}

\end{figure}

We test the dynamical stability of the bright and dark-in-bright solitons of figures \ref{fig2}(e) and \ref{fig3}(a), respectively,
with $N=1000, a=100a_0,$ and $s=40$ by real-time simulation with the pre-calculated stationary state. First we consider the stability of the bright soliton.
During real-time evolution   of the bright soliton of  figure \ref{fig2}(e)
the scattering length was changed from $a=100a_0$ to $99a_0$ at $t=0$. The subsequent evolution of the {root-mean-square (rms) sizes $ x_{\mathrm{rms}}, y_{\mathrm{rms}}, z_{\mathrm{rms}}$ is shown in figure \ref{fig4}(a).  The sustained oscillation of the rms sizes} guarantees the stability of the bright soliton.  
Next we consider the stability of the dark-in-bright soliton.
The  dark-in-bright soliton of figure \ref{fig3}(a) has a notch at  $x=0$. In a time evolution (real-time simulation or experiment) of a normal trapped dark soliton, the position of the notch oscillates around the center (snake instability) and eventually the dark soliton is destroyed \cite{inst2,inst3}. It will be interesting to see if the notch moves away from center at $x=0$ during real-time evolution of the present dark-in-bright soliton. To test this snake instability,      
 we plot the linear  (1D) density along $x$ direction, defined by 
\begin{eqnarray}\label{1dx}
n_{1D}(x) =\int_{-\infty}
^{\infty}dy  \int_{-\infty}
^{\infty}dz |\phi({\bf r})|^2,
%\\   
%&n_{1D}(x) =  \int_{-\infty}
%^{\infty}dz |\phi_{2D}({\boldsymbol \rho})|^2,
\end{eqnarray}  
in figure \ref{fig4}(b)   at different times $t$ 
during real-time evolution after a   change of the scattering length $a$ from 
$100a_0$ to $95a_0$ at $t=0$. Because of the large perturbation, the linear density is found 
to oscillate, however,   maintaining the notch fixed at $x=0$. The rms sizes also  oscillate around a mean value (not shown here) as in the case of the bright soliton shown in figure \ref{fig4}(a). 
This illustrates  clearly that the notch remains in tact for a long time and the dark-in-bright soliton is dynamically stable.

\begin{figure}[!t]

\begin{center}
\includegraphics[width=.49\linewidth,clip]{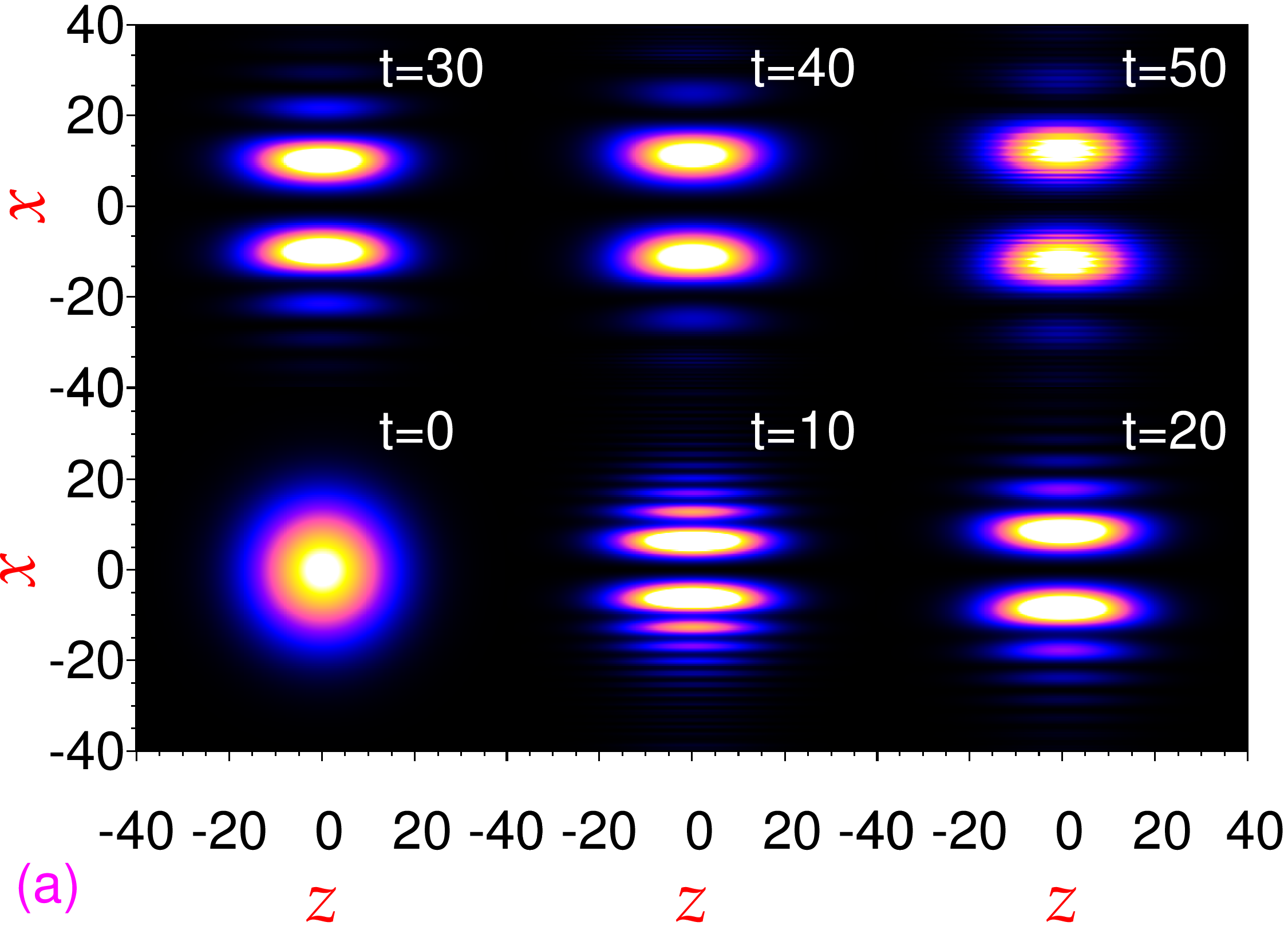}
\includegraphics[width=.49\linewidth,clip]{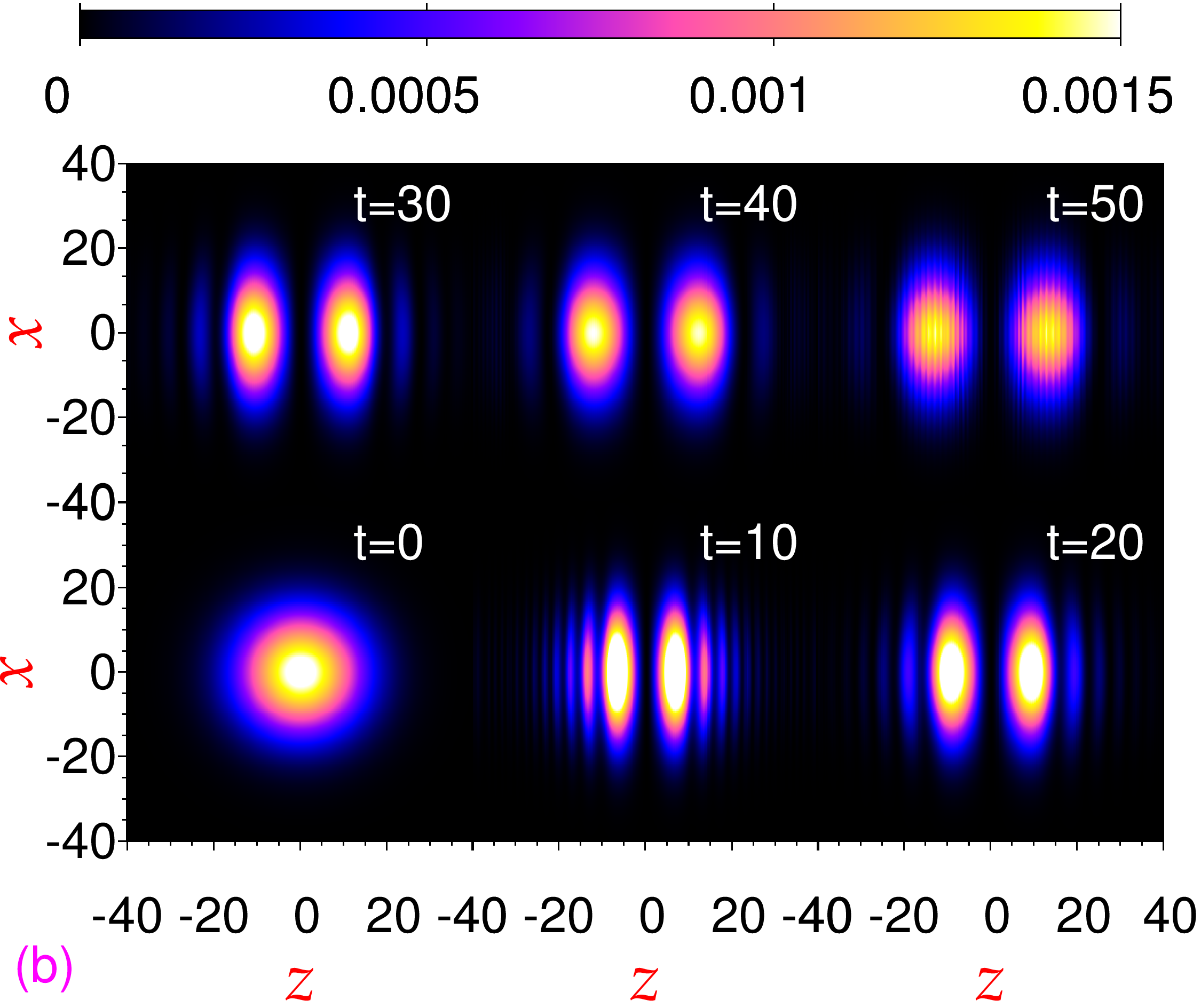}
\caption{(Color online) Creating the quasi-2D dark-in-bright solitons of  (a) figure \ref{fig4}(a) and (b) figure \ref{fig4}(b)  by real-time evolution with parameters $N=1000, a=100a_0, s=40$
starting from  initial  
phase-imprinted Gaussian profiles  $\phi(x,y,z) =- \phi(-x,y,z)$ and 
 $\phi(x,y,z) =- \phi(x,y,-z)$, respectively, at $t=0$.
}\label{fig5} \end{center}

\end{figure}

As the quasi-2D dipolar dark-in-bright solitons are dynamically stable without any snake instability, these can be created by real-time simulation from the following phase-imprinted Gaussian profiles  
\begin{eqnarray}
&&\phi({\bf r})=\phi({\bf r}), \quad x\ge 0;\quad \phi({\bf r})=-\phi({\bf r}), \quad x< 0,\\
&&\phi({\bf r})=\phi({\bf r}), \quad z\ge 0;\quad  \phi({\bf r})=-\phi({\bf r}), \quad z< 0,
 \end{eqnarray}
 respectively, for a notch along $x=0$ and $z=0$, where $\phi({\bf r})$ is the bright soliton (\ref{bright}).
The real-time simulation is started with this wave function. 
In an actual experiment a homogeneous potential generated by the dipole
potential of a far detuned laser beam is applied on part of the Gaussian profile
($x<0$ or $ z<0$) 
 for a time interval  appropriate  to imprint an extra phase $\pi$ 
on that part of the wave function. In the experiment the Gaussian profile should be the 
quasi-2D bright soliton with the same number of atoms $N$ and scattering length $a$. In the present real-time simulation the analytic Gaussian profile (\ref{bright}) with $w_x=w_z=15, w_y=1$  
is employed with the appropriate parameters: $N=1000, a=100a_0$ and $s=40$. 
During real-time simulation the phase-imprinted Gaussian profile 
 slowly transforms into a quasi-2D
  dipolar dark-in-bright soliton. Although the present real-time  simulation is done with no trap,  in  an experiment a
 weak in-plane trap can be used during generating the quasi-2D dark-in-bright soliton
starting from a quasi-2D bound state. The in-plane trap should
be eventually removed. 
 The result of simulation is displayed in figures \ref{fig5}(a) and (b) for the dark-in-bright solitons with a notch along $x=0$ and $z=0$, respectively, where we show the  contour plot 
of the 2D density $n_{2D}(x,z)$
at different times.  It is illustrated that at large times the density of the two dark-in-bright solitons of figures \ref{fig5}(a) and (b) 
evolves towards that
of the quasi-2D dark-in-bright  solitons of  figures \ref{fig3}(c) and (d), respectively.

\begin{figure}[!t]

\begin{center}
\includegraphics[trim = 0cm 10mm 0cm 180mm, width=.32\linewidth,clip]{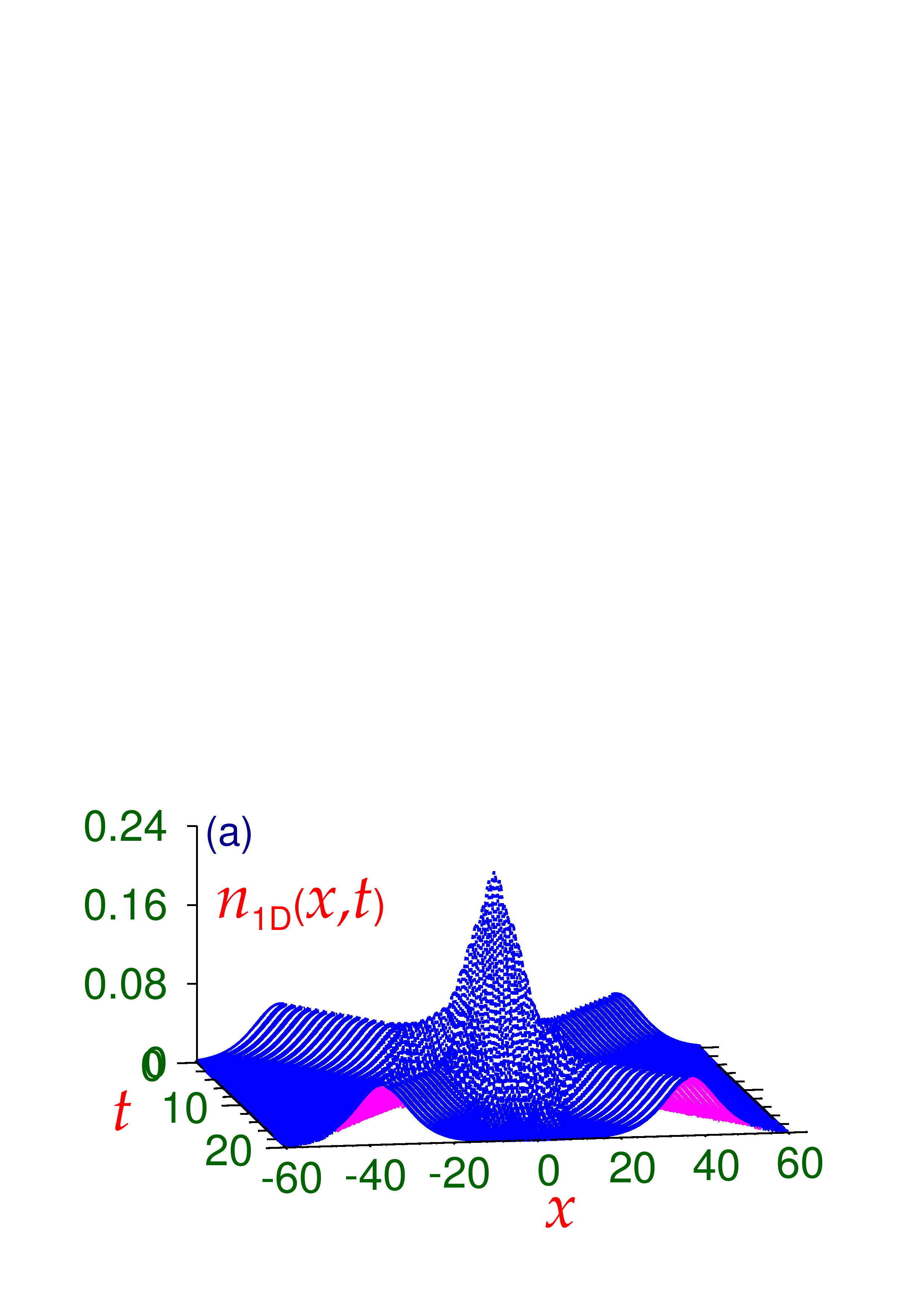}
\includegraphics[width=.32\linewidth,clip]{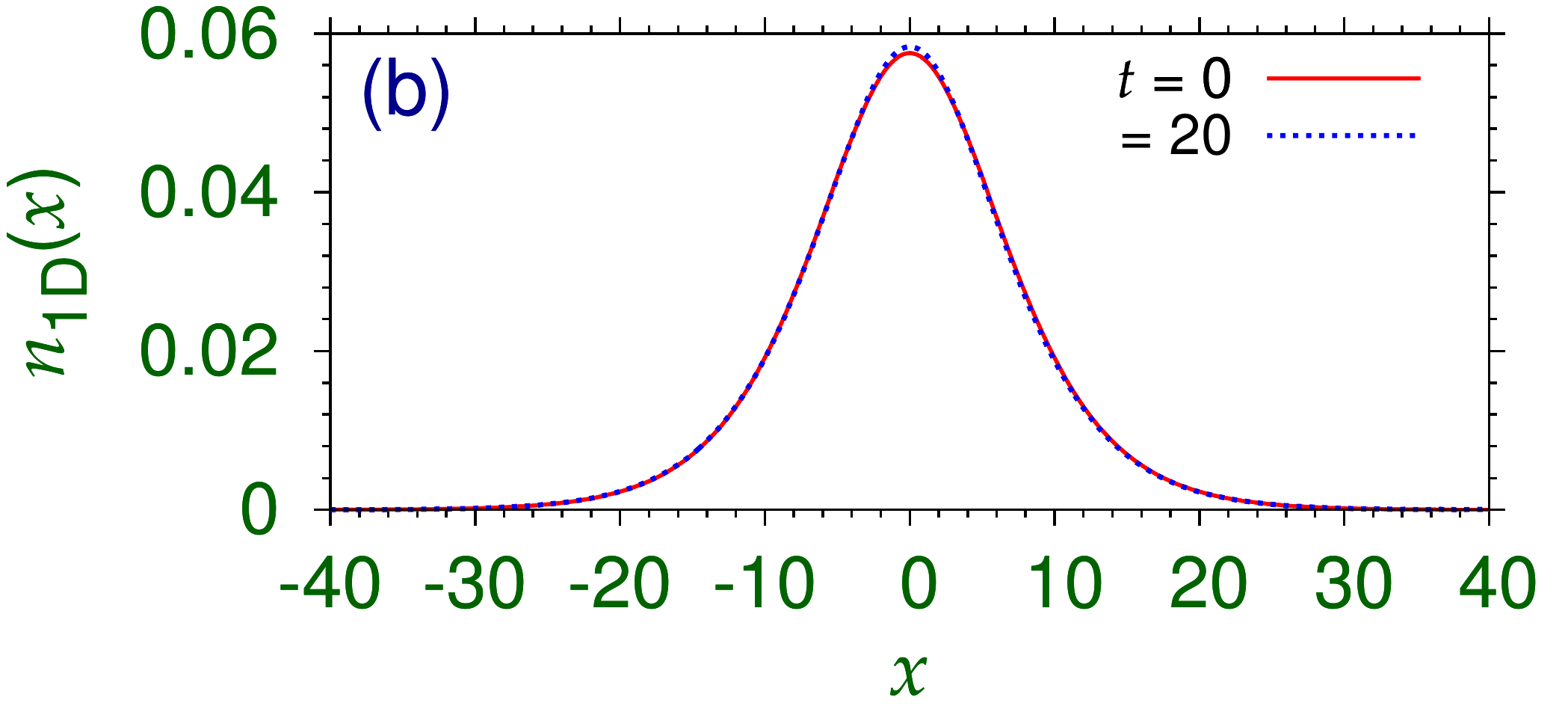}
\includegraphics[width=.32\linewidth,clip]{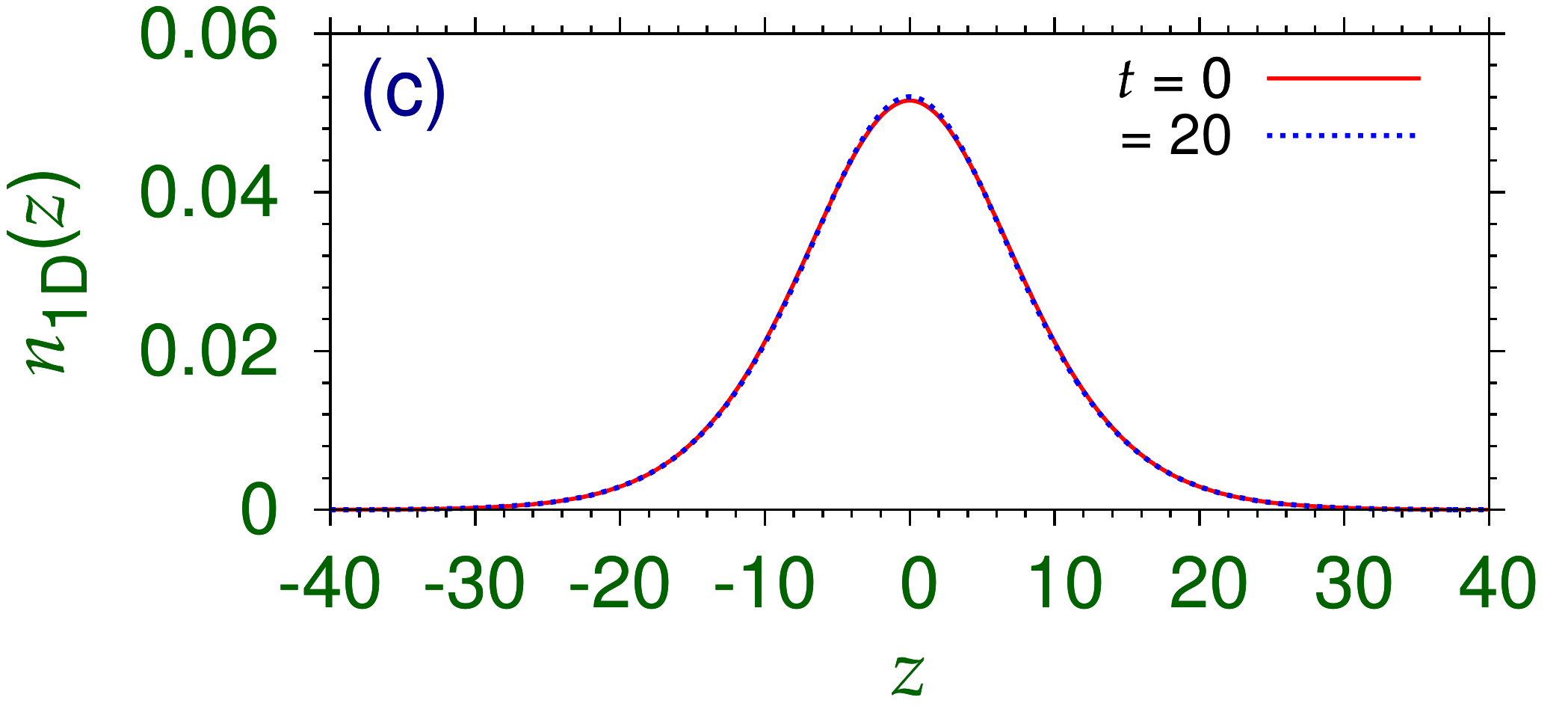}
\caption{(Color online) (a) Frontal  collision dynamics, of two identical 
solitons of figure \ref{fig2}(d) initially placed at $x=\pm 40$ 
and moving  in opposite directions along the $x$ axis each with speed 4, via
a plot of integrated  linear density $n_{1D}(x,t)$
  versus
$x$
and
$t$. (b) Initial ($t=0$) and final ($t=20$) integrated linear densities
  $n_{1D}(x)$ and  $n_{1D}(z)$  of one of the two solitons undergoing 
frontal collision shown in (a).
}\label{fig6} \end{center}

\end{figure}

To demonstate further the stability and robustness of the solitons we consider the 
frontal collision of two solitons of figure \ref{fig2}(d), with $N=100, a=40a_0$ and 
$s=40$,
moving along the $x$ 
axis in opposite directions with a medium velocity.
{ The initial number  of atoms $N$ $(=100)$ in each soliton is chosen in such a way that  during collision of two solitons  
there is no collapse  as found in a recent investigation \cite{hulet}.   This will require 
a value of $N_{\mathrm crit}> 200$ in figure \ref{fig1} for $a=40a_0$.}
The constant velocity of about  4 was attributed to two   solitons  at $x= \pm 40 $ by   multiplying the soliton wave functions  by factors $\exp(\pm i20x)$.
Although the solitons are capable of moving in the $x-z$ plane these phase factors 
will give the solitons a non-zero velocity along the $x$ axis. 
 The initial  wave functions were pre-calculated by imaginary-time propagation. 
In figure \ref{fig6}(a) we plot 
the integrated 1D density $n_{1D}(x,t)$ of the dynamics of collision obtained by a real-time 
simulation of the GP equation.  In figure \ref{fig6}(b) and (c) we plot the integrated 
initial ($t=0$) and final ($t=20$) 
1D
densities $n_{1D}(x)$ and  $n_{1D}(z)$, respectively,  of one of the two solitons undergoing collision.
After the collision the solitons emerge quasi unchanged 
demonstrating  the solitonic nature. 
{For the choosen set of parameters no oscillation in the shape of emerging solitons was found as noted in Ref.  \cite{hulet}.}

\section{Conclusion}

We demonstrated  the possibility of creating stable quasi-2D bright and dark-in-bright solitons in dipolar BEC
mobile in a plane ($x-z$) containing the  polarization direction ($z$). The quasi-2D geometry is obtained by an OL trap in the $y$ direction. The result and finding are illustrated 
by a numerical solution of  the time-dependent   
mean-field GP equation in 3D  with realistic values of contact and  dipolar interactions of $^{164}$Dy    atoms.    
   The solitons are found to execute sustained breathing oscillation 
upon a small perturbation. The dark-in-bright soliton can be created in real-time simulation 
starting from an initial bright soliton, where an extra phase of $\pi$ is introduced in the appropriate half. No dynamical instability is found in the real-time evolution of the dark-in-bright soliton after introducing a small perturbation.
By real-time simulation we demonstrate the elastic nature of the frontal collision between 
two quasi-2D solitons. 
 The results and conclusions  of the present paper can be  tested in experiments with present-day know-how and 
technology  and should lead to interesting future investigations.

%\acknowledgments  
The work has been supported by the 
Funda\c c\~ao de Amparo \`a Pesquisa do Estado de S\~ao Paulo Project 2012/00451-0
(Brazil) and the Conselho Nacional de Desenvolvimento Cient\'ifico 
e Technol\'ogico Project 303280/2014-0 (Brazil).

\end{document}